\documentclass[fleqn]{article}
\usepackage{cite}
\usepackage{graphicx}
\graphicspath{{.}}
\DeclareGraphicsExtensions{.eps,.ps,.eps.gz,.ps.gz}
\newcommand{\ftwo}{\mbox{$F_2(x,Q^2) \;$}}

\newcommand{\ftwopom}{\mbox{$F_2^\mathcal{P}(x,Q^2)$}}
\newcommand{\ftworeg}{\mbox{$F_2^\mathcal{R}(x,Q^2)$}}

\newcommand{\gev}{{\rm Ge}\kern-1.pt{\rm V}}
\newcommand{\gevsq}{\mbox{$\mathrm{{\rm Ge}\kern-1.pt{\rm V}}^2$}}
\newcommand{\gamstar}{\mbox{$\gamma^*$}}

\newcommand{\kev}{{\rm ke}\kern-1.pt{\rm V}}

\newcommand{\mev}{{\rm Me}\kern-1.pt{\rm V}}

\newcommand{\mosq}{\mbox{$m_0^2$}}

\newcommand{\sigmatot}{\mbox{$\sigma_{\gamma p}^{tot}$}}
\newcommand{\sigmastot}{\mbox{$\sigma_{\gamma^{\ast} p}^{tot}$}}

\newcommand{\htwo}{^2\kern-1.pt{\rm H}}
\newcommand{\hetwo}{^2\kern-1.pt{\rm He}}
\newcommand{\hethree}{^3\kern-1.pt{\rm He}}
\newcommand{\hefour}{^4\kern-1.pt{\rm He}}
\newcommand{\beseven}{^7\kern-1.pt{\rm Be}}
\newcommand{\liseven}{^7\kern-1.pt{\rm Li}}
\newcommand{\beight}{^8\kern-1.pt{\rm B}}
\newcommand{\beeight}{^8\kern-1.pt{\rm Be}}
\newcommand{\cl}{^{37}\kern-1.pt{\rm Cl}}
\newcommand{\ar}{^{37}\kern-1.pt{\rm Ar}}
\newcommand{\ga}{^{71}\kern-1.pt{\rm Ga}}
\newcommand{\ger}{^{71}\kern-1.pt{\rm Ge}}
\def\lsim{\mathrel{\rlap{\lower4pt\hbox{\hskip1pt$\sim$}}
    \raise2pt\hbox{$<$}}} 
\def\gsim{\mathrel{\rlap{\lower4pt\hbox{\hskip1pt$\sim$}}
    \raise2pt\hbox{$>$}}} 
\begin{document}
%
\thispagestyle{empty}
\begin{center}
\begin{LARGE}
{\bf Total Cross Section Measurements\\ in $\gamma$p and \gamstar p at HERA}\\*[1cm]
\end{LARGE}
\begin{large}
{\sc Adolf Bornheim}\\*[3mm]
{\it Physikalisches 
Institut, Universit\"at Bonn\\ Nu{\ss}allee 12, 53115 Bonn, Germany\\
E-mail : bornheim@mail.desy.de}\\*[1cm]
\end{large}
\end{center}
\noindent
Measurements of the total cross section for real and virtual photons on protons at
center-of-mass energies in the range from 20 GeV to 270 GeV with 
photon virtualities up to 5000 $\rm GeV^2$ are presented. 
For real photons this cross section can be described by Regge-motivated models
while for virtual photons perturbative QCD can be applied.
The measurements of the two HERA collider experiments ZEUS and H1 open the possibility 
to investigate the interplay between the two theoretical approaches in the transition 
region as well as the high-energy behavior of the cross sections.
The results of total cross section measurements 
are discussed in the above context.\\*[3cm]
Contribution to the proceedings of the LISHEP International School on High-Energy Physics, Rio de Janeiro, Brazil, 16-20 February 1998
\clearpage
\pagenumbering{arabic}
\setcounter{page}{1}
\section{Introduction}
\label{sec:intro}
\begin{figure}[htbp]
\begin{minipage}[ht]{0.39\textwidth}
\includegraphics[width=6.0cm,height=6.0cm, bb= 280 290 620 700, clip=]{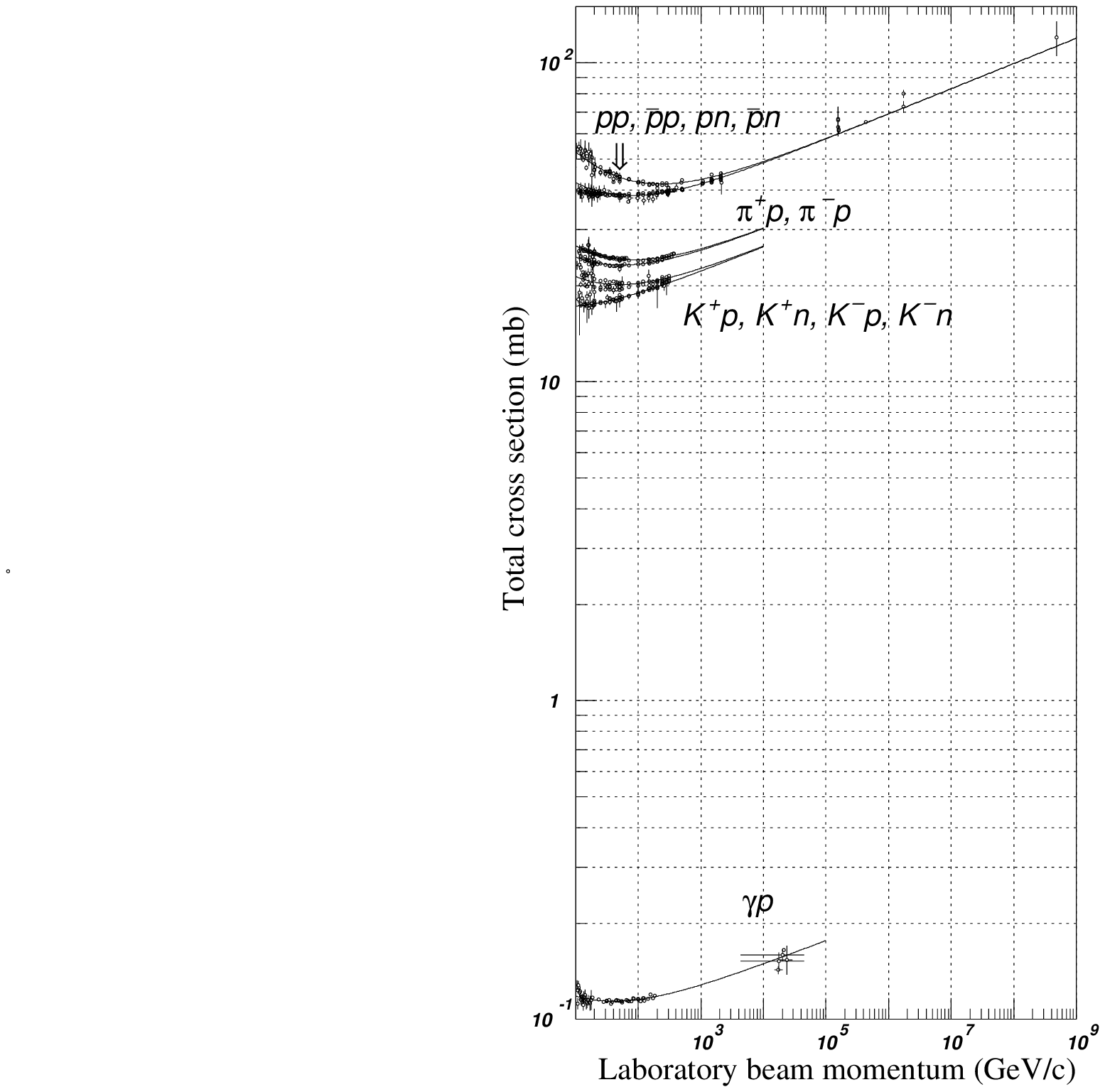}
\caption
{\label{fig:total}
Total cross section as a function of laboratory beam momentum for various hadron-hadron scattering processes.
}
\end{minipage}
\hfill
\begin{minipage}[ht]{0.59\textwidth}
\includegraphics[width=\textwidth, bb= 70 180 550 655, clip=]{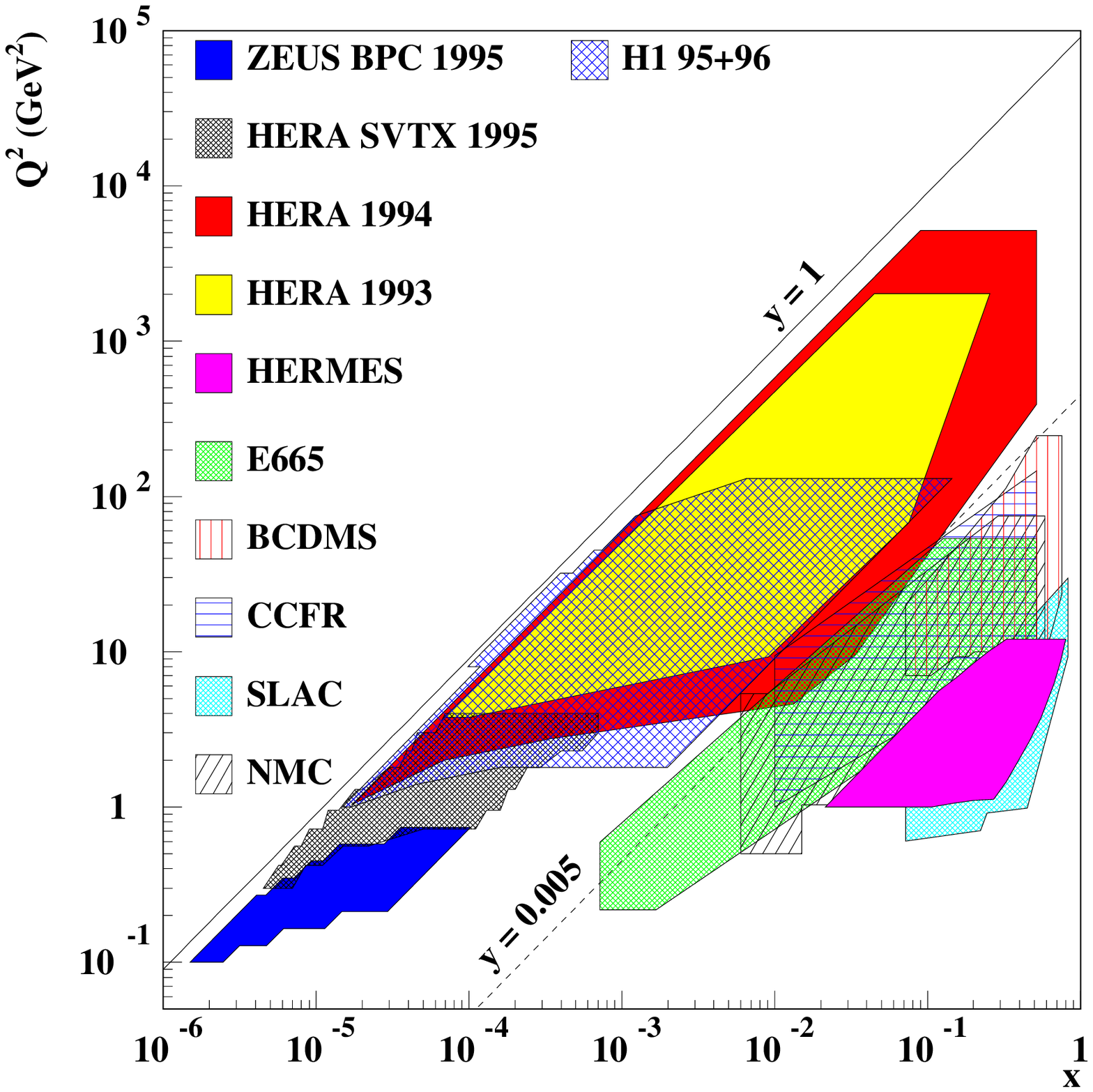}
\caption
{\label{fig:kinplane}
Kinematic range in $Q^2$ and $x$ covered by DIS experiments. 
}
\end{minipage}
\end{figure}

The high-energy behavior of the total cross sections for various hadron-hadron processes
show very similar rises with the center-of-mass energy $W$ as shown in Fig.~\ref{fig:total}. This behavior can be described by the Regge-motivated ansatz \cite{dola92,dola94,cud97,cud96}
\begin{equation}
\label{dleq}
\sigmatot = A_R(W^2)^{\alpha_R-1}+A_P(W^2)^{\alpha_P-1}.
\end{equation}
It is found that this ansatz also describes the photon-proton cross section for real photons which
is referred to as photoproduction (PHP), as will be explained in Sect.~\ref{sec:php} in more detail. 
\par
%
In deep inelastic scattering 
processes (DIS) the application of perturbative (pQCD) is possible due to the presence of a hard scale. In DIS the hard scale is given by the photon virtuality $Q^2$.
This presents the interesting opportunity to compare Regge-motivated models with pQCD calculations and to look for their potential relations.\\
At HERA 27.5 \gev positrons collide with $820 \; \gev$ protons, resulting in a center-of-mass energy
of $300 \; \gev$. 
The kinematic range covered by the experiments including data up to 1996 is shown in Fig.~\ref{fig:kinplane}.
As follows from Eq.~\ref{yandw} the range in $y$ is equivalent to a range in $W$ from  $20 \; \gev$ to $270 \; \gev$. 
The wide coverage in these variables offers a unique possibility to investigate the high-energy
behavior of the total cross section for photon-hadron interactions as a function of
photon virtuality $Q^2$.
This report provides a review of recent results and a discussion of the global features of
 the total cross sections.
\clearpage
\section{Kinematics of Electron-Proton Scattering}
\label{sec:kine}
The cross section of the electron-proton scattering process can be expressed in any
two of the four variables \\
\begin{minipage}[t]{0.06\textwidth}
$Q^2$ \\ \\
$x$ \\
$y$ \\ \\
$W$ \\
\end{minipage}
\hfill
\begin{minipage}[t]{0.93\textwidth}
the negative square of the four momentum transfer between electron and quark,\\ 
the fraction of the proton momentum carried by the struck quark, \\
the relative energy transfer from the electron to the proton in the proton rest system and \\
the center of mass energy of the photon-proton system.\\
\end{minipage}
\par
\vspace*{-0.5cm}
\begin{figure}[ht]
\begin{minipage}[t]{0.4\textwidth}
\includegraphics[width=1.5\textwidth, bb= 150 620 550 800, clip=]{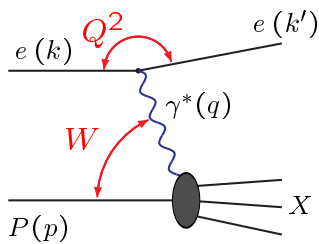}
\end{minipage}
\hfill
\begin{minipage}[ht]{0.6\textwidth}
\vspace*{-4.0cm}
\caption
{\label{fig:diag}
Schematic view of the high-energy scattering process $e p \rightarrow e X$ , illustrating the variables $W$ and $Q^2$.
}
\end{minipage}
\end{figure}
They are related as follows:
\begin{equation}
\label{yandw}
y=\frac{Q^2}{sx} \\
W^2=m_{\rm p}^2+Q^2 \left( \frac{1}{x}-1 \right) \\
$ yielding $ \hspace*{0.3cm} W^2 \approx ys 
\end{equation} 
with $s$ being the squared center-of-mass energy of the electron-proton interaction and $m_{\rm p}$ the proton mass.\\ 
In terms of cross sections $(\sigma_{\rm L})$ and $(\sigma_{\rm T})$ for longitudinally and transversely polarized photons the 
electron-proton cross section can be written as 
\begin{equation}
\label{sigmay}
\frac{{\rm d^2}\sigma_{\rm ep}}{{\rm d}y {\rm d}Q^2} = \frac{\alpha}{2\pi}\frac{1}{Q^2} \left[  \left( \frac{1+(1-y)^2}{y} - \frac{2(1-y)}{y}\frac{Q^2_{\rm min}}{Q^2} \right) \sigma_{\rm T}+\frac{2(1-y)}{y}\sigma_{\rm L} \right],
\end{equation}
where terms of the order $\frac{m_{\rm p}}{s}$ have been neglected, and where
\begin{equation}
Q^2_{\rm min} \approx m^2_{\rm e}\frac{y^2}{(1-y)}.
\end{equation}
For the total cross section measurements in PHP to be presented below $y$ is between 
0.4 and 0.6 for the ZEUS experiment and between 0.3 and 0.7 for the H1 experiment. 
Therefore $Q^2_{\rm min}$ is of the order of $10^{-8} \; \rm GeV^2$ .\\ 
DIS measurements presented here are restricted to the region $y < 0.82$. 
Therefore the $Q^2_{\rm min}$ term in Eq.~\ref{sigmay} is ignored in DIS. 
Writing the cross section as a function of $x$ rather than $y$ yields
\begin{equation}  
\frac{{\rm d}^2\sigma_{\rm ep}}{{\rm d}x {\rm d}Q^2} =\frac{\alpha}{2\pi}\frac{(1-x)}{xQ^2}Y^+(\sigma_T+\epsilon\sigma_L)
\end{equation}
with
\begin{equation}
Y^+ = 1+(1-y)^2 \hspace*{0.6cm} $and$ \hspace*{0.6cm} \epsilon=\frac{2(1-y)}{Y^+}.
\end{equation}
In leading order the cross sections are related to the structure functions by
\begin{equation}  
F_{\rm 2} = \frac{Q^2}{4\pi^2\alpha}(1-x)(\sigma_{\rm T}+\sigma_{\rm L}) \hspace*{1.0cm} F_{\rm L} = \frac{Q^2}{4\pi^2\alpha}(1-x)\sigma_{\rm L}
\end{equation}
were $F_2$ is the sum over the quark and antiquark densities in the proton.
Only one-photon exchange is considered here, so we neglect the structure function $F_3$.
The evolution of the proton structure function $F_2$ is calculable in pQCD
once the distributions $q (\overline{q})$ are known at a starting scale $Q^2_0$. 
$F_L$ can be calculated in pQCD as well and is a small correction to the total
cross section in the kinematic range considered here.
We can therefore write
\begin{equation}  
\sigmastot(x,Q^2) \approx \frac{4 \pi^2 \alpha}{Q^2} \ftwo
\end{equation}    
This means the measurement of the proton structure function $F_2$ as a function of $x$
is equivalent to measuring the total $\gamstar p$ cross section as a function of $x$. Making 
use of relation \ref{yandw} this can be transformed into the total cross section as 
a function of W, the center-of-mass energy of the photon-proton interaction.\\
In PHP $\frac{\sigma_{\rm L}}{\sigma_{\rm T}} << 1$. Taking this into account together and integrating
over the accepted range in $Q^2$ Eq. \ref{sigmay} reduces to
\begin{equation}  
\frac{{\rm d}\sigma_{\rm ep}}{{\rm d}y} =\frac{\alpha}{2\pi}\frac{Y^+}{y}\Big[\ln\Big(\frac{Q^2_{\rm max}}{Q^2_{\rm min}}\Big)-\epsilon\Big(1- \frac{Q^2_{\rm min}}{Q^2_{\rm max}}\Big)\Big]\sigmatot(W_{\gamma p})
\end{equation}    
with 
\begin{equation}  
W_{\gamma p} = \sqrt{4E_\gamma E_p} \\ E_\gamma =yE_e.
\end{equation}
Ignoring the second term in the square brackets yields the Weizs\"acker-Williams approximation \cite{wwa2} which, for the present experimental situation, overestimates the flux of photons by approximately $7 \%$ \cite{burrow}.\\
The above formulae show how the total cross section in PHP and DIS are related and
how the latter is related to the parton densities, whose evolution in
$Q^2$ is calculable in pQCD.
It should be pointed out that the distinction between DIS and PHP in the following is 
based on experimental constraints rather than different theoretical approaches.\\ 
%
%
%
%
\section{Experimental Techniques}
\label{sec:experiment}
Details of the experimental techniques of structure function measurements from the 
collaborations H1 and ZEUS can be found elsewhere \cite{h1f294,zeusf294}
as well as details on the structure function measurements at low $Q^2$~\cite{bpc95,h1lowq295} and the 
total cross section measurements in PHP \cite{zeustot92,zeustot94,h1tot95}.\\
For medium and high $Q^2$ DIS events both the scattered electron and the hadronic system are 
measured in the main detector, 
for low $Q^2$ DIS the electron is measured in a small calorimeter close to the beam pipe outside 
the main detector $3 \rm m$ away from the interaction region while the hadronic system is measured
 in the main detector.
For PHP the electron is detected in a small calorimeter about $35 \rm m$ away
from the nominal interaction point, the hadronic system is measured in the main detector.\\ 
For DIS the electron and the hadronic particles produced in the interaction 
are used for event classification and for measuring the kinematic variables.
Due to this combination of information from the electron side and the hadron side 
no single systematic uncertainty dominates over the whole phase space.
Instead one can roughly
summarize the situation as follows: at high $W$ systematic uncertainties arise primarily from
the electron side, while at low $W$ they come mainly from the hadron side.
The combined statistical and systematic uncertainties are typically a few percent, reaching
more than 10 percent at the very edge of the accessible phase space.\\
In the PHP case, the electrons and the hadrons are used to trigger the event as well as for 
event classification, but the kinematics is calculated from the electron only. 
Therefore the systematic
error is dominated by two contributions: first, the uncertainty in the acceptance of the 
electron calorimeter, secondly, the uncertainty in the MC simulation of 
the topology of the hadronic final state produced, which introduces an uncertainty in the 
acceptance of the main detector. 
In contrast to the
situation in DIS the acceptance varies strongly with the event topology in PHP.
While the determination of the acceptance of the electron calorimeter is a
technical aspect, the uncertainty in the modelling of the hadronic final state 
also reflects our limited knowledge of the underlying physics process. 
The investigation of these processes 
is an important field of study at HERA.\\  
It should be stressed here that from an experimental point of view there is no major difference between 
DIS and PHP. In the following, events with the scattered electron in the main detector or the 
beam pipe calorimeter (ZEUS) are referred to as DIS. This corresponds to a lower limit in $Q^2$ of $0.1 \; \rm GeV^2$
while the $W$ range reaches from 20 GeV to 270 GeV.
For PHP events the scattered electron is measured in a calorimeter close to the beam pipe 35 m away from the interaction point for the ZEUS experiment 
and 33 m for the H1 experiment. This corresponds
to a lower limit on $Q^2$ of the order of $10^{-8} \; \rm GeV^2$. The mean $W$ is determined by the position of the electron 
calorimeter and the energy range of the scattered electrons which reaches from 15.2 GeV to
18.2 GeV for the ZEUS measurement and 8.3 GeV to 19.25 GeV for the H1 experiment. 
The total cross sections in PHP are determined for a single value of W.\\ 
%
%
%
%
%
%
%
%
%
%
\clearpage
\section{Results from Deep Inelastic Scattering}
\label{sec:dis}
\begin{figure}[!ht]
\begin{minipage}[t]{0.47\textwidth}
\includegraphics[width=\textwidth, bb= 60 125 540 720, clip=]{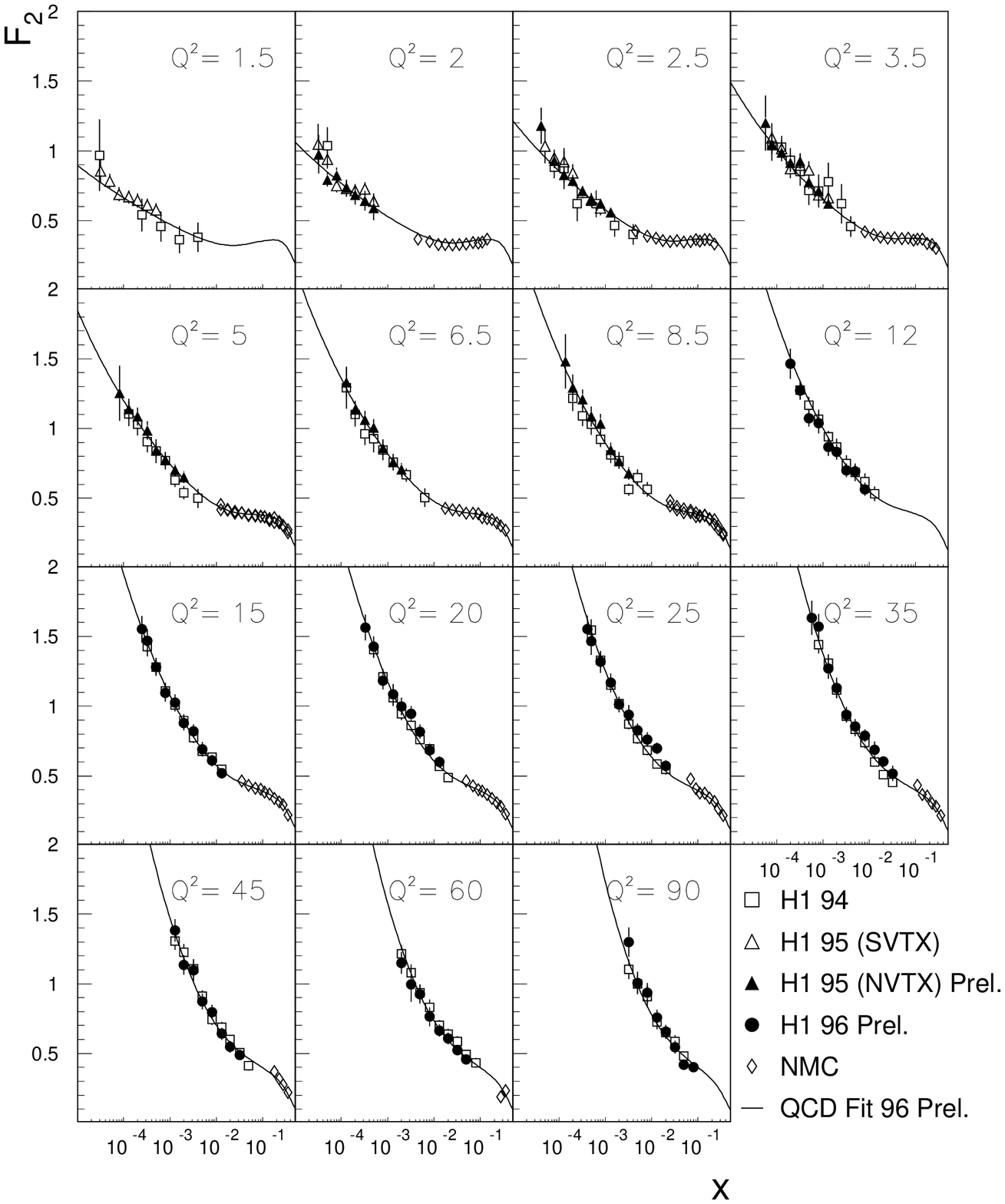}
\caption
{
\label{fig:f2lq2h1jeru.ps}
Preliminary results of the proton structure function \ftwo at low $Q^2$ .
The line shows the result of a QCD fit to the data.
}
%
\end{minipage}
\hfill
\begin{minipage}[t]{0.47\textwidth}
\vspace*{-7.0cm}
\includegraphics[width=\textwidth,height=6.5cm ,bb= 80 330 514 680, clip=]{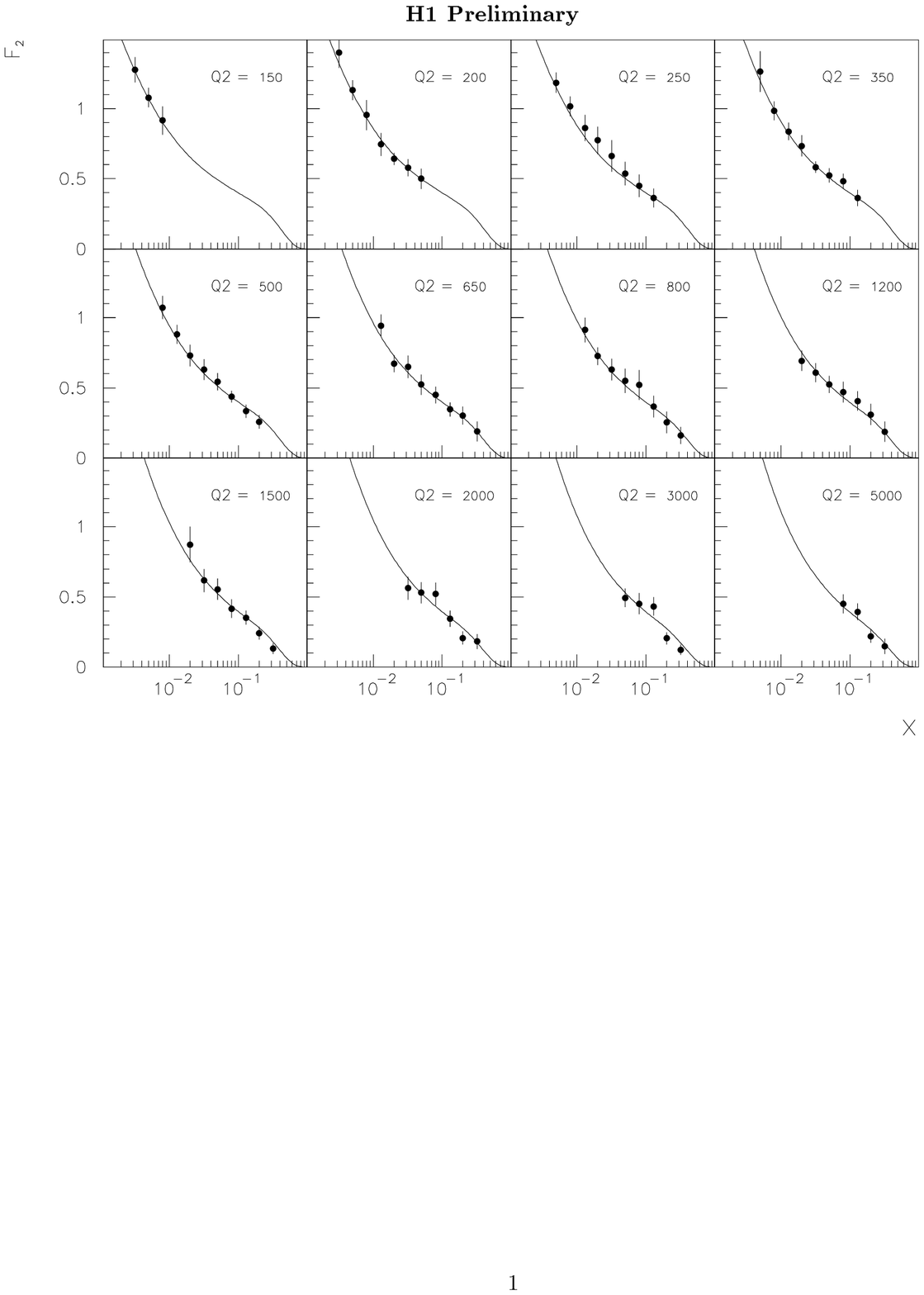}
\caption
{\label{fig:f2hq2h1jeru}
Preliminary results on the proton structure function $F_2$ at high $Q^2$.
The line shows the QCD fit to the low $Q^2$ data, extrapolated to high $Q^2$.
}
\end{minipage}
\end{figure}

\begin{figure}[ht]
\begin{minipage}[ht]{0.4\textwidth}
\vspace*{3.5cm}
\includegraphics[width=\textwidth, bb= 30 30 550 480, clip=]{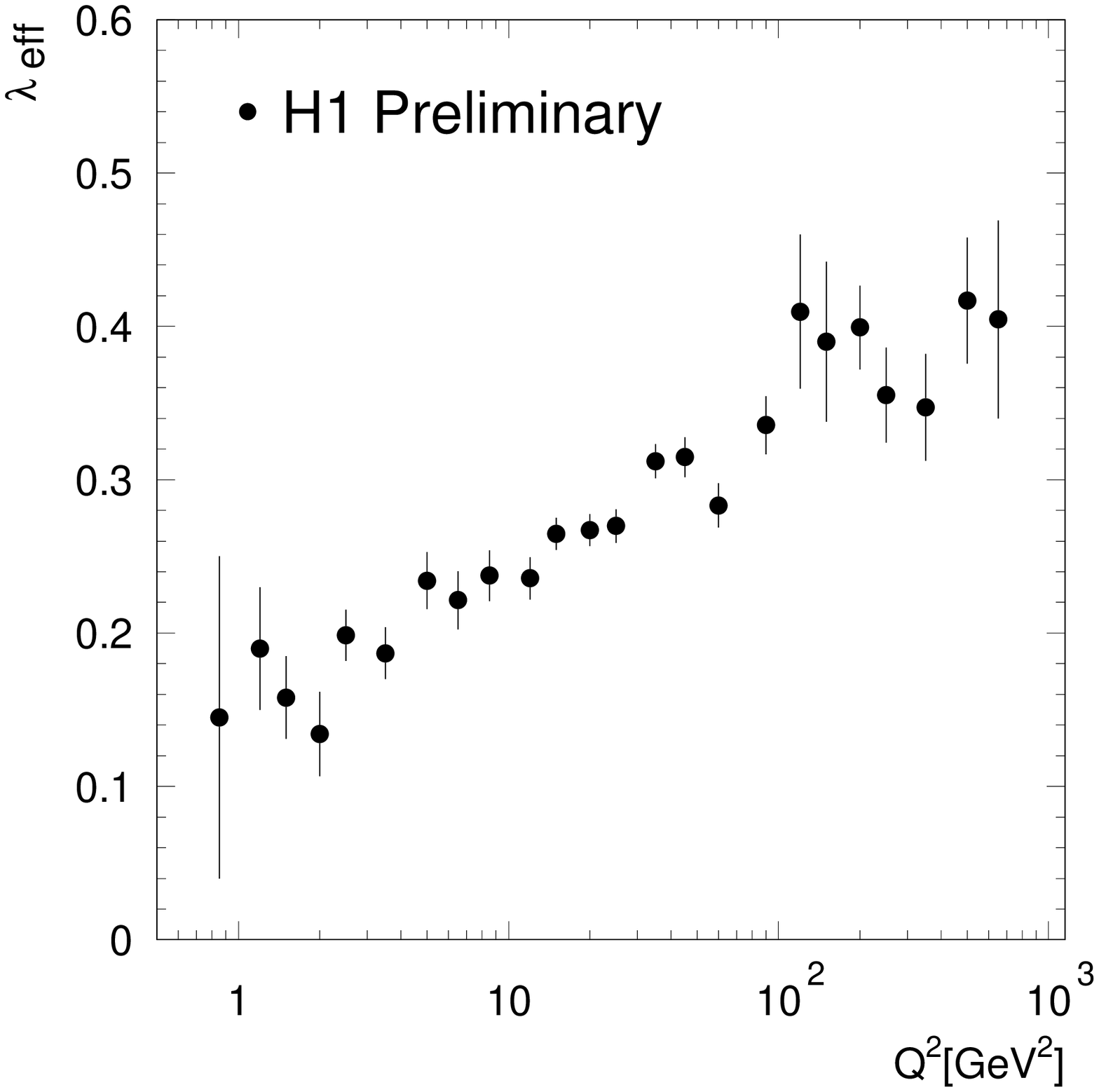}
\caption
{\label{fig:lambda}
The slope $\lambda_{\rm eff}$ of $F_2$ as a function of $Q^2$ .
}
\end{minipage}
\hfill
\begin{minipage}[ht]{0.57\textwidth}
\includegraphics[width=\textwidth, bb= 40 125 550 700, clip=]{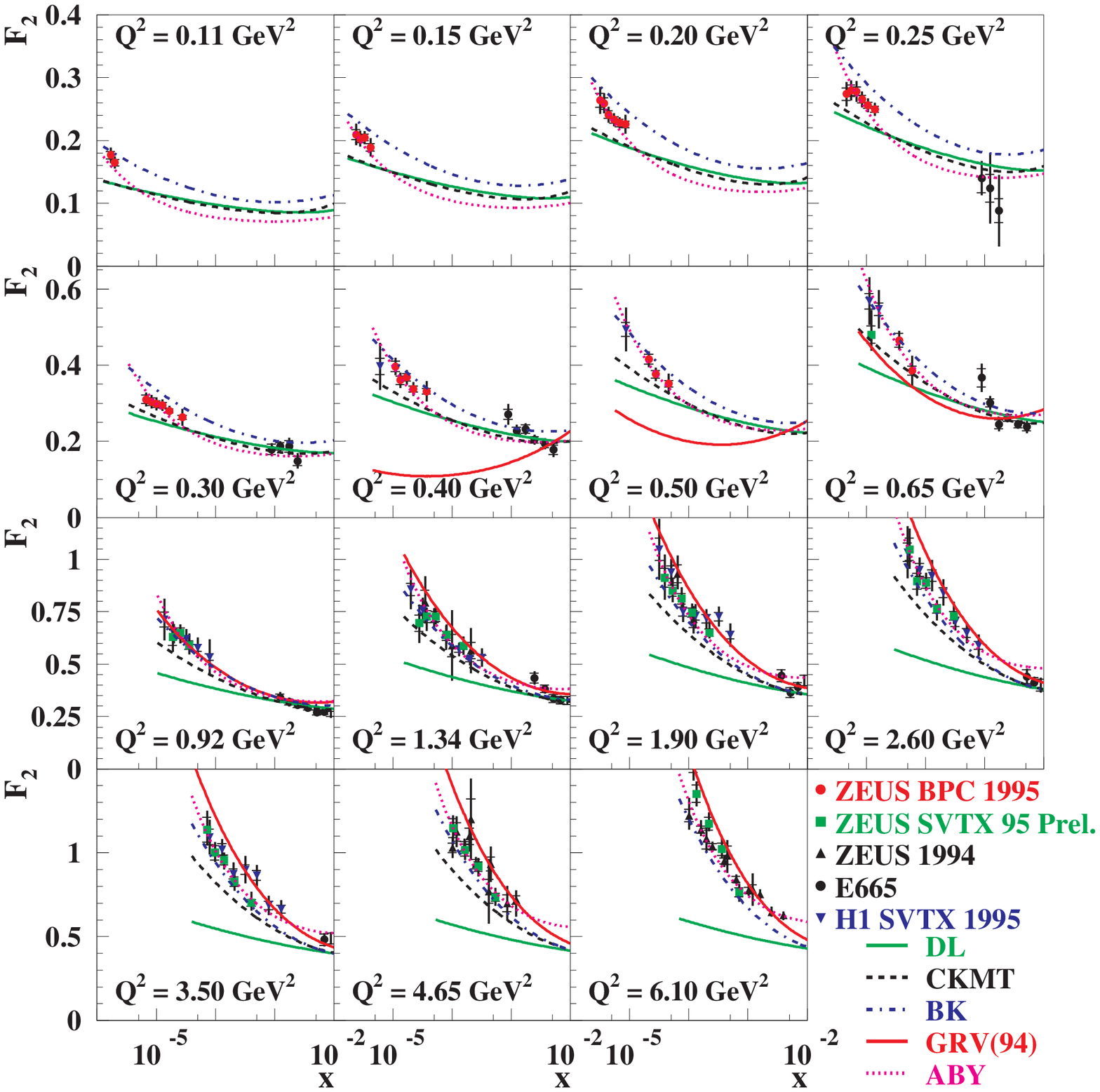}
\caption
{\label{fig:bpc}
Results on the proton structure function \ftwo \hspace{0.1cm} at low $Q^2$.
}
\end{minipage}
\end{figure}

The measurement of the proton structure function $F_2$ is one of the major achievements of HERA.
Several measurements have been performed taking advantage of the increasing luminosity and
improved detectors and reconstruction methods \cite{zeusf292,zeusf293,zeuslowq294,zeusf294,bpc95,
h1f292,h1f293,h1f294,h1lowq295}.
The results shown here are preliminary results from the H1 collaboration.
The low $Q^2$ data shown in Fig.~\ref{fig:f2lq2h1jeru.ps} is from the running period 1996 
\cite{h1jerulowq2}, compared to previously published data. 
The high $Q^2$ data shown in Fig.~\ref{fig:f2lq2h1jeru.ps} is from the running periods 
1996 and 1997 \cite{h1jeruhighq2}. 
A QCD fit has been performed to the low $Q^2$ data and then extrapolated to the high $Q^2$ regime.
The very good agreement found is an impressive validation of pQCD in this kinematic range.\\
\vspace*{-0.4cm}
\vspace*{-0.2cm}
%
%
%
\newline
As the focus is on the high-energy behavior of the total cross section
it is instructive to inspect the slope of $F_2$ with respect to x for fixed $Q^2$.
This is equivalent to investigating the total cross section as a function 
of $W$ as explained in sect.~\ref{sec:kine}.
In Fig.~\ref{fig:lambda} the slope $\lambda_{\rm eff}$ is shown which is derived by fitting the function
\begin{equation}
\ln (F_2)=a+\lambda_{\rm eff} \ln \left( \frac{1}{x} \right) \hspace*{1.0cm} x < 0.1
\end{equation}
to the measured structure function data from various analyses \cite{h1f294,h1lowq295,h1jerulowq2}.
$F_2 \propto x^{-\lambda}$ implies $ \sigmastot \propto W^{2\lambda}$,
hence the result from this fit can be compared with the value of $\alpha_{P}$
in Eq. \ref{dleq}.\\
\par
\vspace*{-0.5cm}
As can be seen in Fig.~\ref{fig:lambda} the slope of $F_2$ is found to rise with increasing
$Q^2$. For decreasing $Q^2$ it approaches the value which is found in hadron-hadron scattering.
The gap between the DIS measurements and the PHP measurements is covered by structure function
measurements using the ZEUS beam-pipe calorimeter \cite{bpc95}. The results of this measurement 
are shown in Fig.~\ref{fig:bpc}. 
They are compared to recent models which aim to describe the transition from PHP to DIS.
The Donnachie-Landshoff ansatz \cite{dola94} which assumes a power behavior as shown in Eq.~\ref{dleq},
the CKMT model \cite{ckmt} which introduces a $Q^2$ dependence of this power-law rise of 
the cross section with W. The $Q^2$ dependence is fitted using NMC data.
The BK ansatz \cite{bk} which is based on the generalized vector meson dominance model,
the ABY model \cite{aby} which combines NLO-QCD fits and a soft component and 
the $F_2$ calculated from the GRV94 parton densities \cite{grv94} which are tuned 
in such a way that they reproduce
the transition from a slow to a steep rise of the cross section.
None of these approaches is able to describe the cross section in the transition region.
It should be mentioned that the ABY model uses the BPC data to fix its parameters while 
the GRV94 parton densities were tuned even before the HERA94 DIS data was available.
The fact that none of the models can reproduce the cross section in the transition region between
DIS and PHP without explicitly using the measurements to constrain the parametrizations 
indicates that neither pQCD nor Regge-motivated models alone are able to describe the 
underlying physics.  
%
%
%
%
%
\clearpage
\section{Results from Photoproduction}
\label{sec:php} 
\begin{figure}[htbp]
\includegraphics[height=6.8cm,width=0.8\textwidth, bb= 0 185 514 610, clip=]{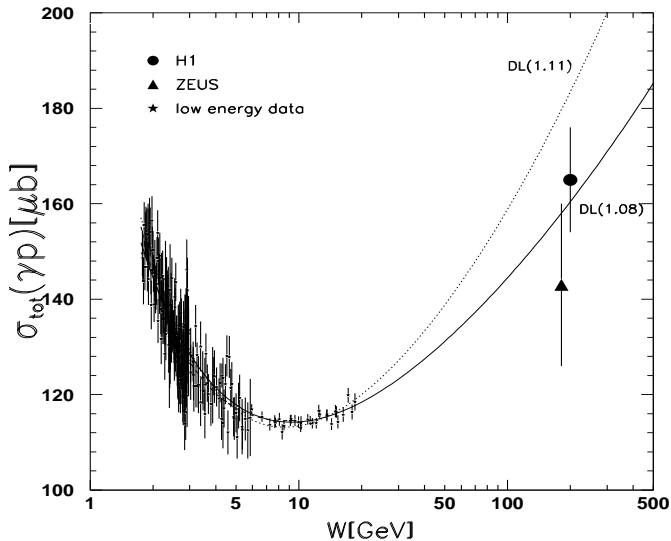}
\caption
{\label{fig:sigmatot}
Results on the total photoproduction cross section
}
\end{figure}

Measurements of the total cross section in PHP have been performed by ZEUS \cite{zeustot94} 
and by H1 \cite{h1tot95}. At present only measurements for a single value of $W$ for each experiment 
are available. 
The results are presented in Fig.~\ref{fig:sigmatot}, 
together with results at lower values of $W$ \cite{caldwell}.
Also shown are curves for the DL parametrization as in Eq.~\ref{dleq} for two values
of $\alpha_{P}$, 1.08 and 1.11, representing the uncertainty on this value from hadron-hadron
data \cite{dola94,cud97}.
\par 
\vspace*{-0.2cm}
In contrast to the situation in DIS, the systematic error in PHP is dominated by two contributions, 
the error arising from the uncertainty in the acceptance of the electron calorimeter and
the uncertainty of the acceptance of the main detector as described in Sect.~\ref{sec:experiment}. 
The exact results and the contributions from the two major sources of systematic uncertainties are
for the H1 experiment\\*[1mm]
\hspace*{0.1cm}$\sigmatot = 165.3 \pm 2.3 (stat.) \pm 10.9 (syst.) \hspace*{0.2cm} \mu \rm b \hspace*{0.3cm}$ at $W = 200 \; \rm GeV$.\\*[1mm] 
From the $ \pm 10.9 \mu \rm b$ systematic error \\
\hspace*{0.5cm}$\pm 8.5 \; \mu \rm b$ are due to uncertainty on the electron calorimeter acceptance and \\
\hspace*{0.5cm}$\pm 5.1 \; \mu \rm b$ are due to the model dependence of the main detector acceptance.\\*[1mm]
The cross section measured by the ZEUS experiment is\\*[1mm]
\hspace*{0.1cm}$\sigmatot = 143 \pm 4 (stat.) \pm 17 (syst.) \hspace*{0.2cm} \mu \rm b \hspace*{0.3cm} $ at $W = 180 \; \rm GeV$.\\*[1mm] 
From the $ \pm 17 \mu \rm b$ systematic error \\
\hspace*{0.5cm}$\pm 13 \; \mu \rm b$ are due to uncertainty on the electron calorimeter acceptance and\\
\hspace*{0.5cm}$\pm 10 \; \mu \rm b$ are due to the model dependence of the main detector acceptance.\\*[1mm]
The results are in agreement with the DL ansatz for hadron-hadron data, but the magnitude of the uncertainties
do not allow a determination of $\alpha_{P}$ with an accuracy comparable to that from the hadronic cross sections.
\section{Behavior of the Total Cross Section}
\label{sec:allm}
\begin{figure}[htbp]
\begin{minipage}[ht]{0.52\textwidth}
\includegraphics[width=1.1\textwidth,height=8.5cm, bb= 30 105 530 700, clip=]{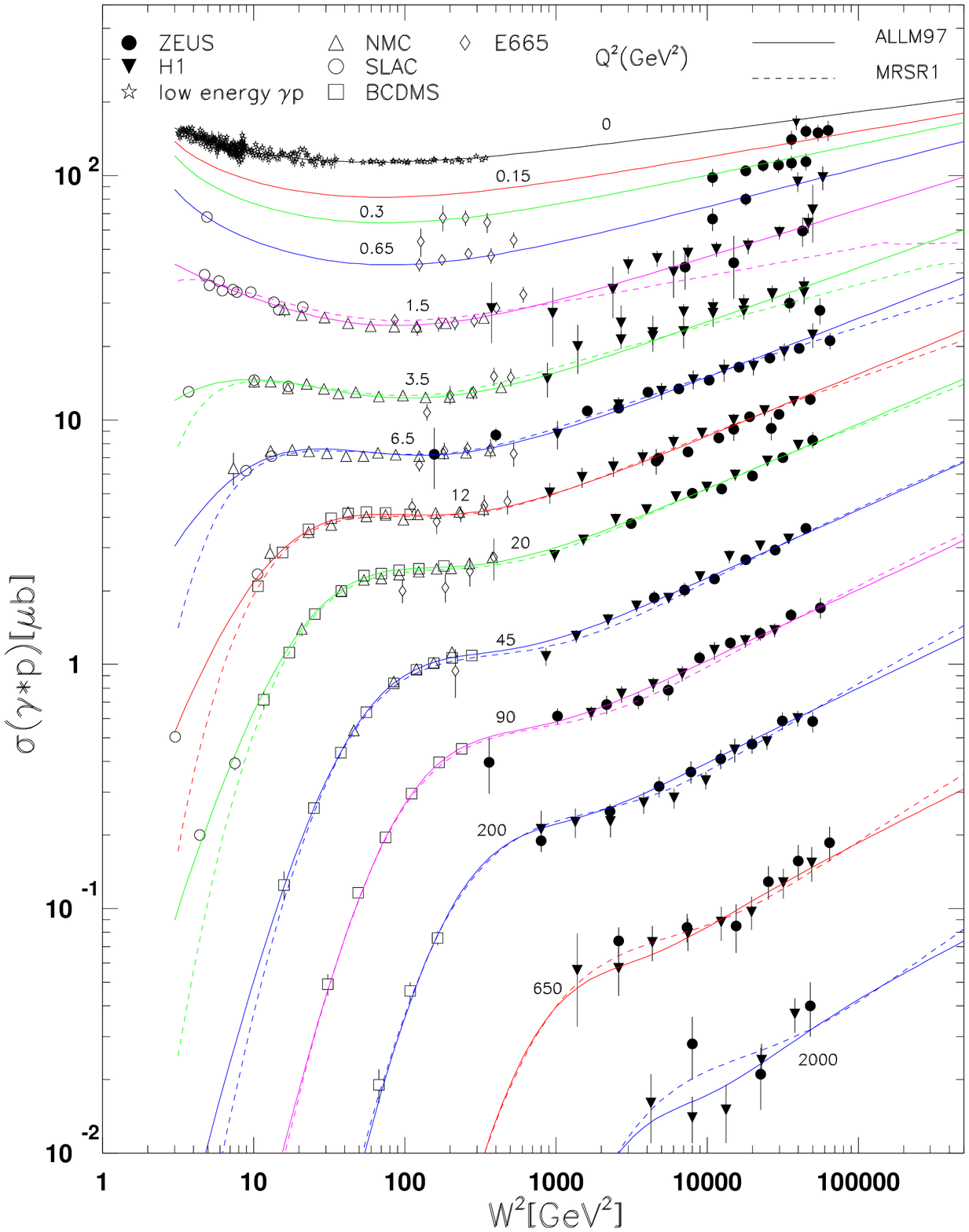}
\caption
{\label{fig:allmfit}
Measured cross sections compared to the ALLM97 fit. Also shown the cross sections calculated from
a recent parton density set (MRSR1). 
}
\end{minipage}
%
\hfill
\begin{minipage}[ht]{0.42\textwidth}
\vspace*{-0.5cm}
\includegraphics[width=1.05\textwidth,height=5.0cm, bb= 30 280 540 700, clip=]{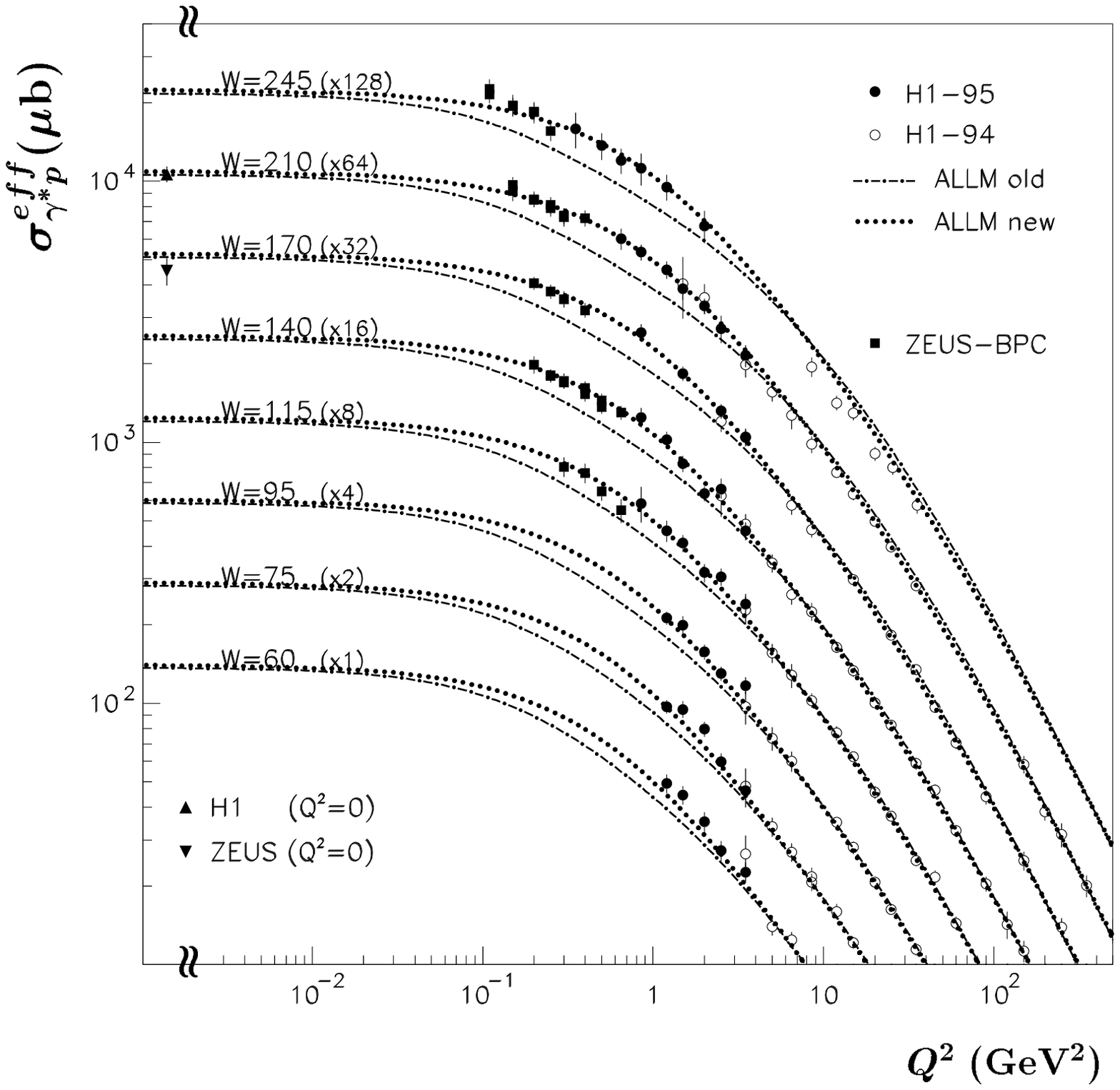}
\caption
{\label{fig:allmlqsq}
The fit at low $Q^2$.
}
\includegraphics[width=1.05\textwidth,height=3.2cm, bb= 0 200 514 650, clip=]{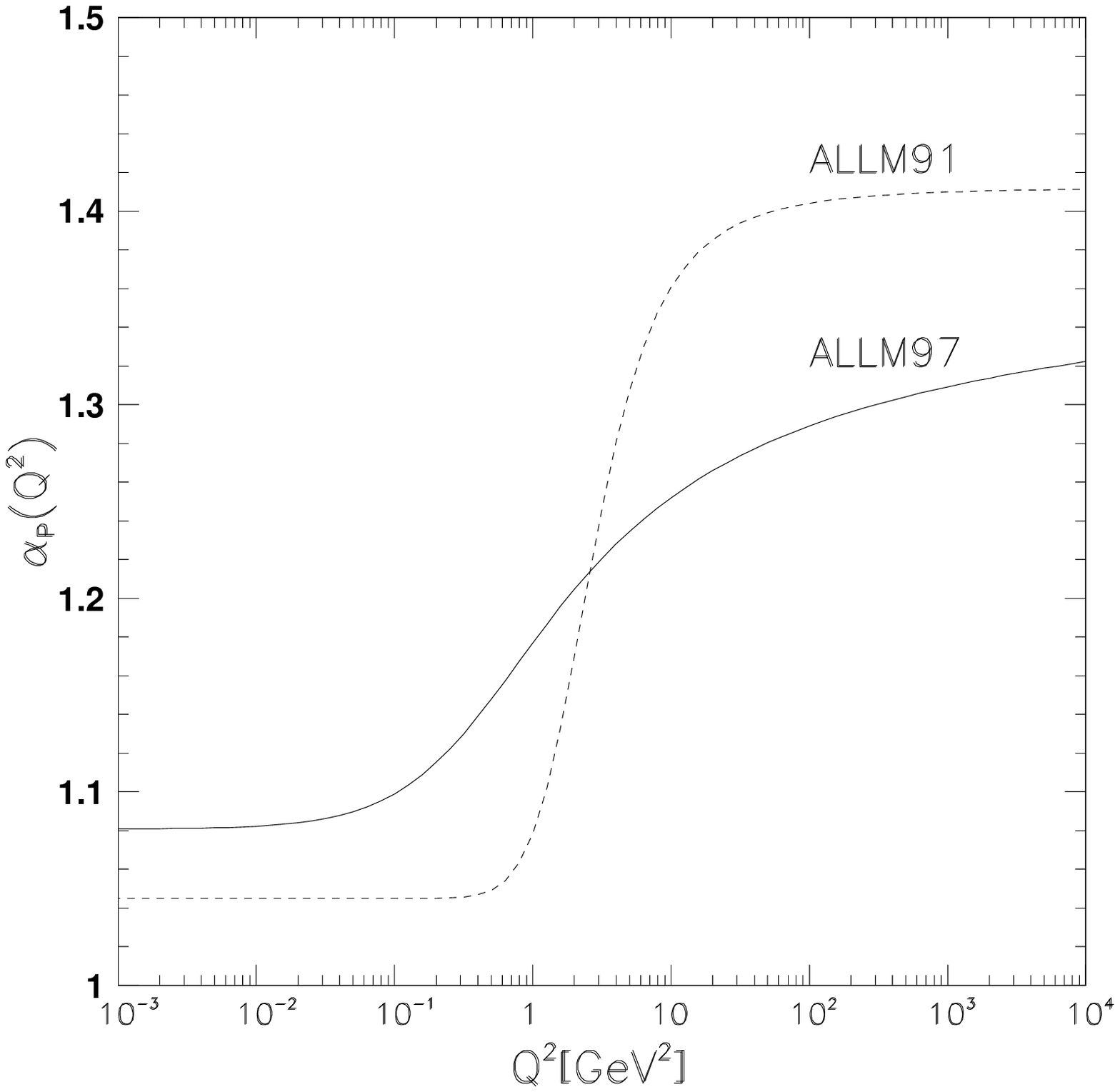}
\caption
{\label{fig:alphapom}
The change of the value of $\alpha_{P}$ due to the inclusion of low $Q^2$ data.
}
\end{minipage}
\end{figure}

The results on cross section measurements presented here cover the $Q^2$ range from 
$10^{-8} \; \rm GeV^2$ to $5000 \; \rm GeV^2$ and the
$W$ range from 20 $\rm GeV$ to $270 \; \rm GeV$.
As neither QCD nor Regge models are able to describe the behavior of the total cross section 
over the complete phase space we use the most recent update of the ALLM \cite{allm,allm97}
parametrization to demonstrate some features of the measured cross section.  
The basic idea of this parametrization is to describe the total cross section with 
a Regge-type ansatz
\begin{equation}
\ftwo=\hspace*{-1mm}\frac{Q^2}{Q^2 + \mosq}\hspace*{-1mm} \left( \ftwopom + \ftworeg \right)
\end{equation}
with two contributing terms, $\ftwopom$ and $\ftworeg$, which 
are assumed to have energy dependences similar to those of the pomeron and reggeon parts 
in the DL ansatz.
A $Q^2$-dependence is introduced in such a way that the behavior is compatible with QCD expectations
at high values of $Q^2$ and on the other hand reproduces the measured cross sections for PHP.
This ansatz is able to describe the data in the whole kinematic region covered by the measurements
(Fig.~\ref{fig:allmfit}).   
\par
\vspace*{0.1cm}
\begin{figure}[htbp]
\hfill
\begin{minipage}[ht]{0.6\textwidth}
\includegraphics[width=\textwidth, bb= 30 170 540 630, clip=]{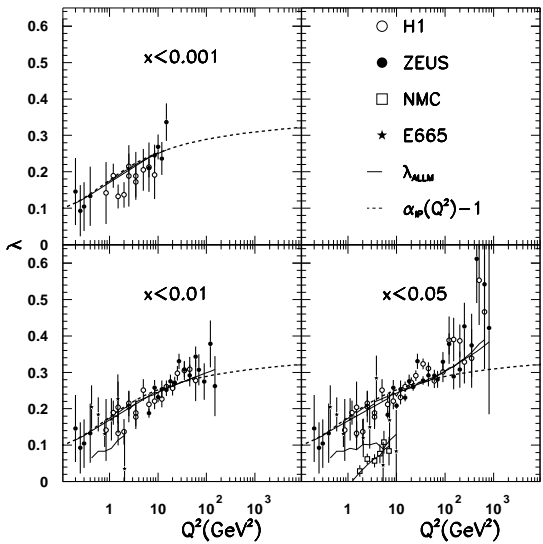}
\end{minipage}
\hfill
\begin{minipage}[ht]{0.39\textwidth}
\caption
{\label{fig:allmlambda}
Comparison of $\lambda$ as determined from the data \cite{h1jerulowq2} and 
the slope calculated from the ALLM97 parametrization. 
}
\end{minipage}
\end{figure}

\noindent
A major difference of the recent update of the parametrization \cite{allm97} compared to 
the initial one \cite{allm} is visible at low $Q^2$ (Fig. \ref{fig:allmlqsq} and 
Fig.~\ref{fig:alphapom}) and is the result of the inclusion of the new data 
\cite{bpc95,h1lowq295} in this region.\\
It is important to note that the intercept for the pomeron-like part $F_2^{P}$ is fixed at
1.08 for $Q^2 = 0 \; \rm GeV^2$, the result of fits to hadron-hadron data, accounting for 
the fact that the errors on the total cross section measurements from HERA at 
$Q^2 = 0 \; \rm GeV^2$ are not small enough to extract this value from HERA data alone.\\ 
This parametrization can be used to characterize the behavior of the total cross section. 
In Fig.~\ref{fig:alphapom} the $Q^2$-dependence of the parameter $\alpha_{P}$ is shown.
This parameter $\alpha_{P}$ as calculated from the 
fit parameters can then be compared to $\lambda_{\rm eff}$ as derived in 
Sect.~\ref{sec:dis} from the data.
The result is shown in Fig.~\ref{fig:allmlambda}. 
We see that there is good agreement between the methods. However, at the highest $Q^2$,
the fit to the data yields somewhat higher values for $\lambda_{\rm eff}$ than what
is calculated from the ALLM97 parameters. This points to systematic uncertainties in determining
this quantity which is of particular interest because its behavior 
as a function of $Q^2$ can also be addressed theoretically by means of pQCD \cite{bfklnlo}. 
A second difference between the old and the new version of the parametrization is the value 
of $\alpha_{P}$ at high $Q^2$, which is significantly lower in the 97 version.
In summary, it can be stated that the change in the behavior of the cross section as a function
of $Q^2$ from flat to a steep rise is a gradual one which develops over a wide range in $Q^2$.
Interpreting the steepness of the rise as an indication whether we are in the soft or in the hard regime
we conclude that the transition between both takes place over a wide range in $Q^2$. 
%
%
%
%
%
%
%
%
\section{Extrapolation of DIS results to ${\mathbf Q^2 \; \mathbf = \; \mathbf 0 \; \rm \mathbf{ GeV^2}}$}
\label{sec:extra}
\begin{figure}[htbpc]
\begin{minipage}[ht]{0.48\textwidth}
\includegraphics[width=\textwidth, bb= 20 60 514 590, clip=]{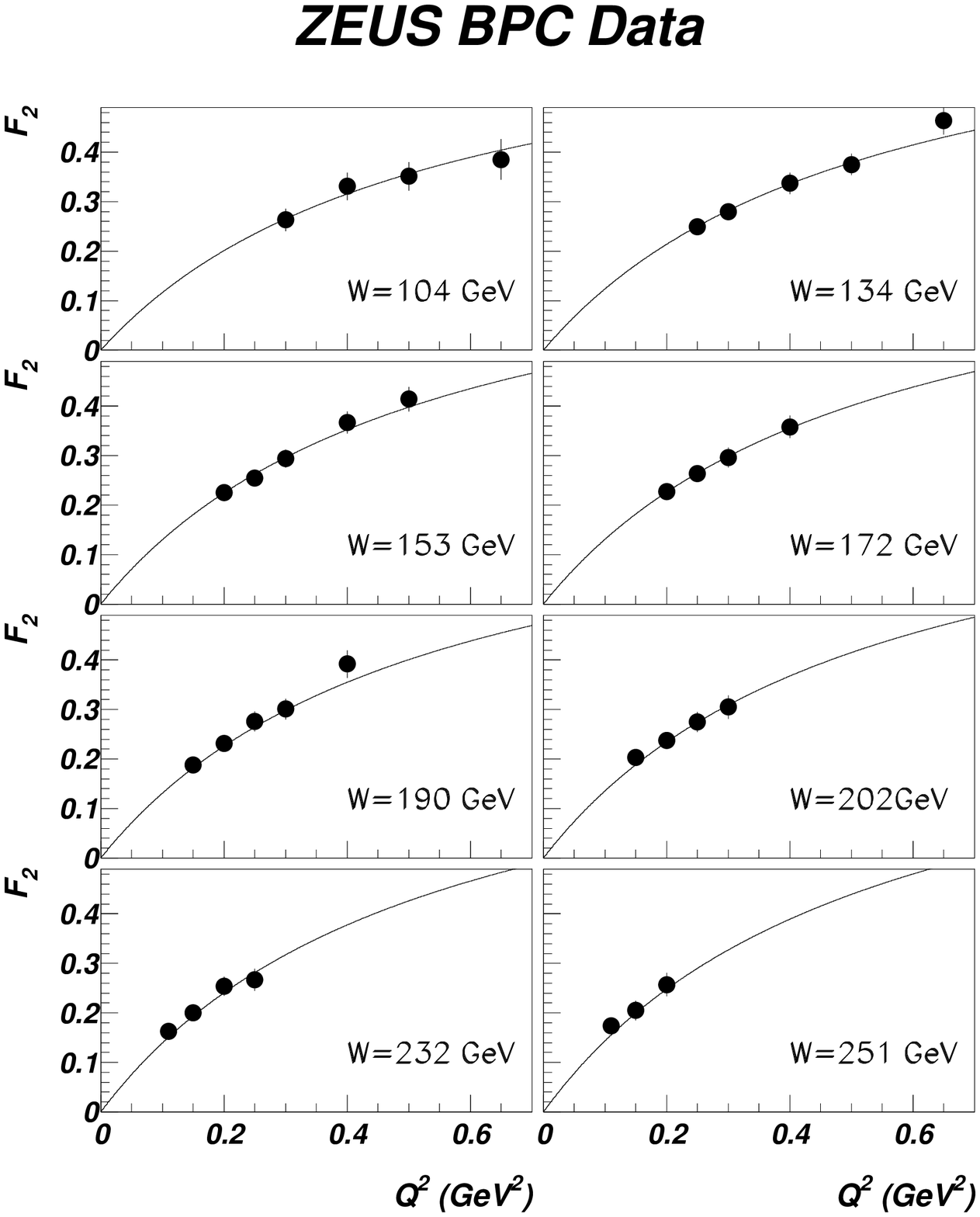}
\caption
{\label{fig:extraolbpc}
Extrapolation of ZEUS BPC results to $Q^2 = 0 \; \rm GeV^2$
}
\end{minipage}
\hfill
\begin{minipage}[ht]{0.48\textwidth}
\includegraphics[width=\textwidth, bb= 30 110 550 720, clip=]{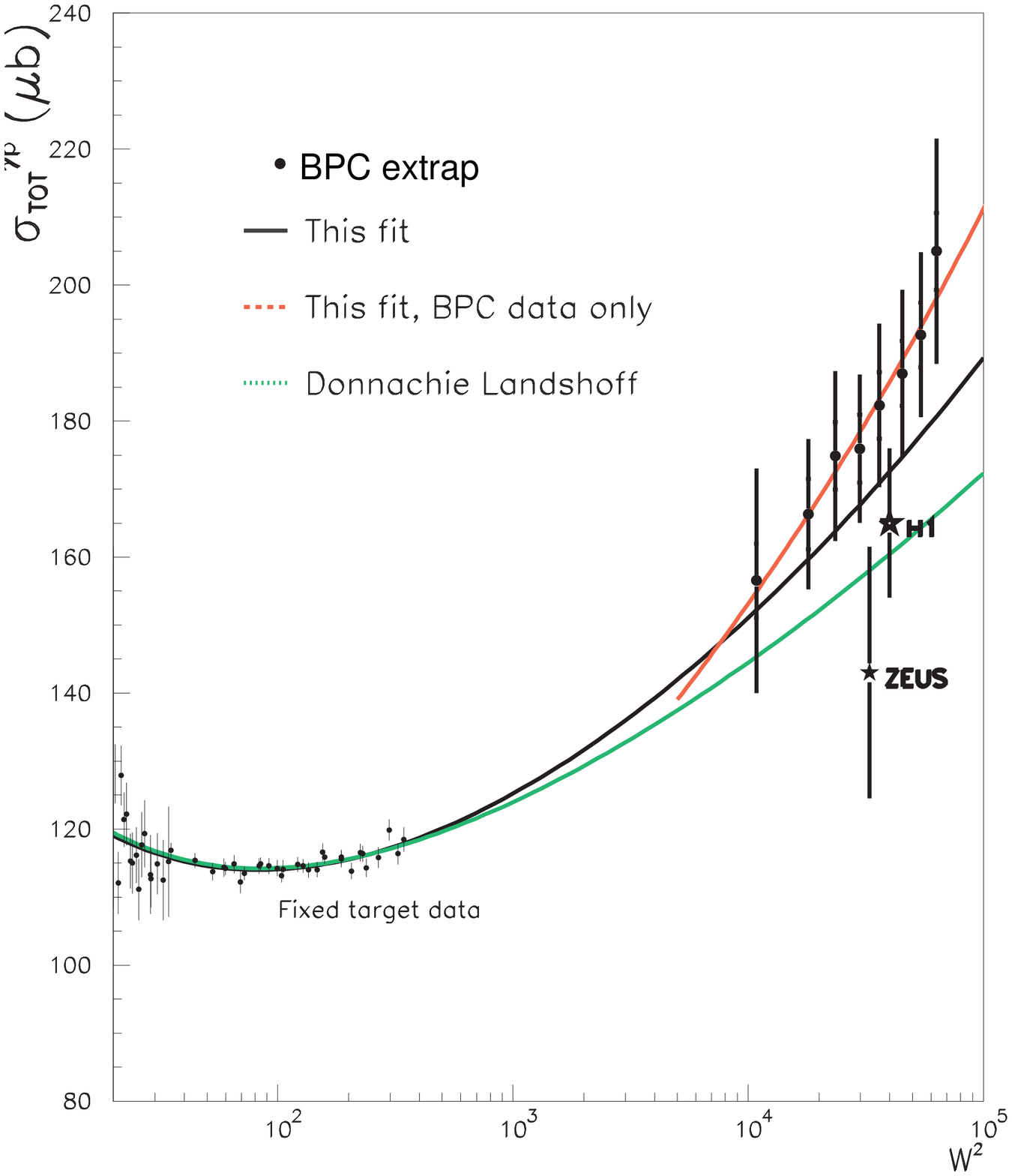}
\caption
{\label{fig:resultextra}
Result of the extrapolation of BPC data to $Q^2 = 0 \; \rm GeV^2$, compared to PHP measurements.
}
\end{minipage}
\end{figure}

As was shown in the previous sections the DIS results allow the determination 
of the dependence of the total cross section on 
$W$ and $Q^2$ and its comparison to pQCD calculations.
On the other hand the results on total cross section measurements in PHP 
allow comparisons with hadron-hadron data and Regge-based models.
To gain more insight into this, an attempt was made to extrapolate 
the DIS data to $Q^2 = 0 \; \rm GeV^2$ .
It should be stressed that the errors on the $F_2$ values and 
hence the cross section values in single bins are not
necessarily much smaller than the error on the total cross section in PHP.  
However, in DIS we benefit from the fact that the large $W$ range covered by
the experiments constrains the rise of the cross section while in PHP we 
have only two measurements at very similar values of W.
As pQCD-calculations do not work at $Q^2 = 0 \; \rm GeV^2$ we have to use an appropriate model 
to do this extrapolation. 
In the example shown here a generalized vector dominance model is used.
The asymptotic behavior for $Q^2 \rightarrow 0 \; \rm GeV^2$ is
\begin{equation}
\sigmastot = \frac{M^2}{Q^2+M^2} \; \sigmatot.
\end{equation}
The fit shown in Fig. \ref{fig:extraolbpc} yields $M= (0.73 \pm 0.03) \rm \; GeV$ \cite{zeuslowq2fit}.
\par
The result of the extrapolation is shown in Fig.~\ref{fig:resultextra}.
The error bars shown include the systematic and statistical uncertainties of the 
$F_2$ measurement and the contribution to the error due to the assumption about $F_L$.
They do not include effects due to model dependences of the extrapolation of 
the cross section to $Q^2 = 0 \; \rm GeV^2$.
More details on this can be found in \cite{tickner,surrow}.
The result of the extrapolation is in agreement with the direct measurements.
However, the error is not significantly smaller than for the direct measurement
and model dependences are not yet included.
Fitting a parametrization like Eq.~\ref{dleq} results in $\alpha_{P} = 1.107$
if the low energy data is included and
\begin{equation}
\alpha_{P} = 1.14 \pm 0.04  \hspace*{0.5cm} $with $ \alpha_{R} $ set to $ 0
\end{equation}
if the fit is applied to the extrapolated BPC data only.
It should be stressed that the fit to the extrapolated data alone is feasible because
they cover a wide range in $W$ while the direct measurement of the total PHP cross section does not.
Both values are in agreement with hadron-hadron data \cite{dola94,cud97}. 

\section{Conclusions and Outlook}
\label{sec:conclusions}
The total cross section in PHP shows a moderate rise with $W$ compatible with 
the rise of the hadron-hadron cross section.
In DIS this rise becomes steeper with increasing $Q^2$. A phenomenological fit yields a 
logarithmic slope of $\approx 1.1$ at $Q^2 = 1 \; \gevsq$
which rises to $\approx 1.4$ at $Q^2 = 5000 \; \gevsq$.    
The change of the slope takes place gradually over the entire $Q^2$ range.
The behavior of the structure function \ftwo from which the total cross section is calculated 
in DIS is in good agreement with pQCD calculations for $Q^2 \ge 1 \; \rm GeV^2$. 
Extrapolations of DIS results to $Q^2 = 0 \; \rm GeV^2$ are in agreement with direct measurements
and with results from hadron-hadron data.

\section{Acknowledgements}
\label{sec:thanks}
I would like to thank the organizers of lishep98 for a very interesting workshop.
I am grateful to DESY for financial support.
It is a particular pleasure for me to acknowledge the outstanding effort of many colleagues in the
H1 and ZEUS collaborations who contributed to the results which are presented here. 
Finally I would like to thank J. Crittenden, E. Hilger, U. Katz, A. Doyle and A. Quadt for careful reading of the manuscript. 
\addcontentsline{toc}{chapter}{References}
\bibliographystyle{/afs/desy.de/user/b/bornheim/tex/talk/procrio/zeusstylem}
\bibliography{lishep98,procrefs,zeuspubs,h1pubs,schriftrefs}

\end{document}